\documentclass[conference]{IEEEtran}
\IEEEoverridecommandlockouts
\usepackage{cite}
\usepackage{amsmath,amssymb,amsfonts}
\usepackage{graphicx}
\usepackage{textcomp}
\usepackage{xcolor}
\usepackage{float}
\usepackage{booktabs}
\usepackage{multirow}
\usepackage{url}
\usepackage[hidelinks]{hyperref}
\usepackage{caption}
\usepackage{tikz}
\usepackage{cuted}

\usetikzlibrary{shapes.geometric, arrows.meta, positioning}
\def\BibTeX{{\rm B\kern-.05em{\sc i\kern-.025em b}\kern-.08em
    T\kern-.1667em\lower.7ex\hbox{E}\kern-.125emX}}
\begin{document}

\title{Phoenix: Safe GitHub Issue Resolution via Multi-Agent LLMs}

\author{%
\IEEEauthorblockN{%
  kipngeno koech\IEEEauthorrefmark{1}\IEEEauthorrefmark{2}\IEEEauthorrefmark{3},\;
  Muhammad Adam\IEEEauthorrefmark{1}\IEEEauthorrefmark{2}\IEEEauthorrefmark{3},\;
  Baimam Boukar Jean Jacques\IEEEauthorrefmark{1}\IEEEauthorrefmark{2}\IEEEauthorrefmark{3},\;
  Joao Barros\IEEEauthorrefmark{1}\IEEEauthorrefmark{2}\IEEEauthorrefmark{3}\IEEEauthorrefmark{4}%
}
\IEEEauthorblockA{\IEEEauthorrefmark{1}Department of Electrical and Computer Engineering}
\IEEEauthorblockA{\IEEEauthorrefmark{2}Carnegie Mellon University Africa, Kigali, Rwanda}
\IEEEauthorblockA{\IEEEauthorrefmark{3}Carnegie Mellon University, Pittsburgh, USA}
\IEEEauthorblockA{\IEEEauthorrefmark{4}Heinz College of Information Systems and Public Policy}
\IEEEauthorblockA{\{bkoech, madam2, bbaimamb, jbarros\}@andrew.cmu.edu}%
}

\maketitle

\begin{abstract}
We present Phoenix, a multi-agent LLM system that resolves GitHub issues from triage through pull-request creation, combining seven layered safety controls with a baseline-aware test evaluation strategy. Phoenix decomposes the work across six specialized agents. Planner, reproducer, coder, tester, failure analyst and Pull Request (PR) agent, all coordinated by a label-based GitHub webhook state machine. 
Every change is checked against a baseline test run before a pull request is opened. On a 24-instance slice of SWE-bench Lite 
%
run on the production webhook path, Phoenix oracle-resolves 75\% of instances with no pass-to-pass regressions on successful runs; this curated slice is not directly comparable to full-split leaderboard results, and we discuss the limits of the comparison. A complementary pilot on 42 real issues across 14 repositories yields 100\% correctness preservation ($CP$; mean 122\,s on the hard tier). Manual inspection shows that about half of the resulting pull requests are well-targeted fixes. The other half place code at incorrect paths, a planner localization limitation we are addressing with retrieval. We also report the deployment failure modes (WAF filtering, token expiry, permission boundaries, flaky CI) that motivated each safety mechanism.
\end{abstract}


\begin{strip}
    \centering
    \includegraphics[width=1\textwidth]{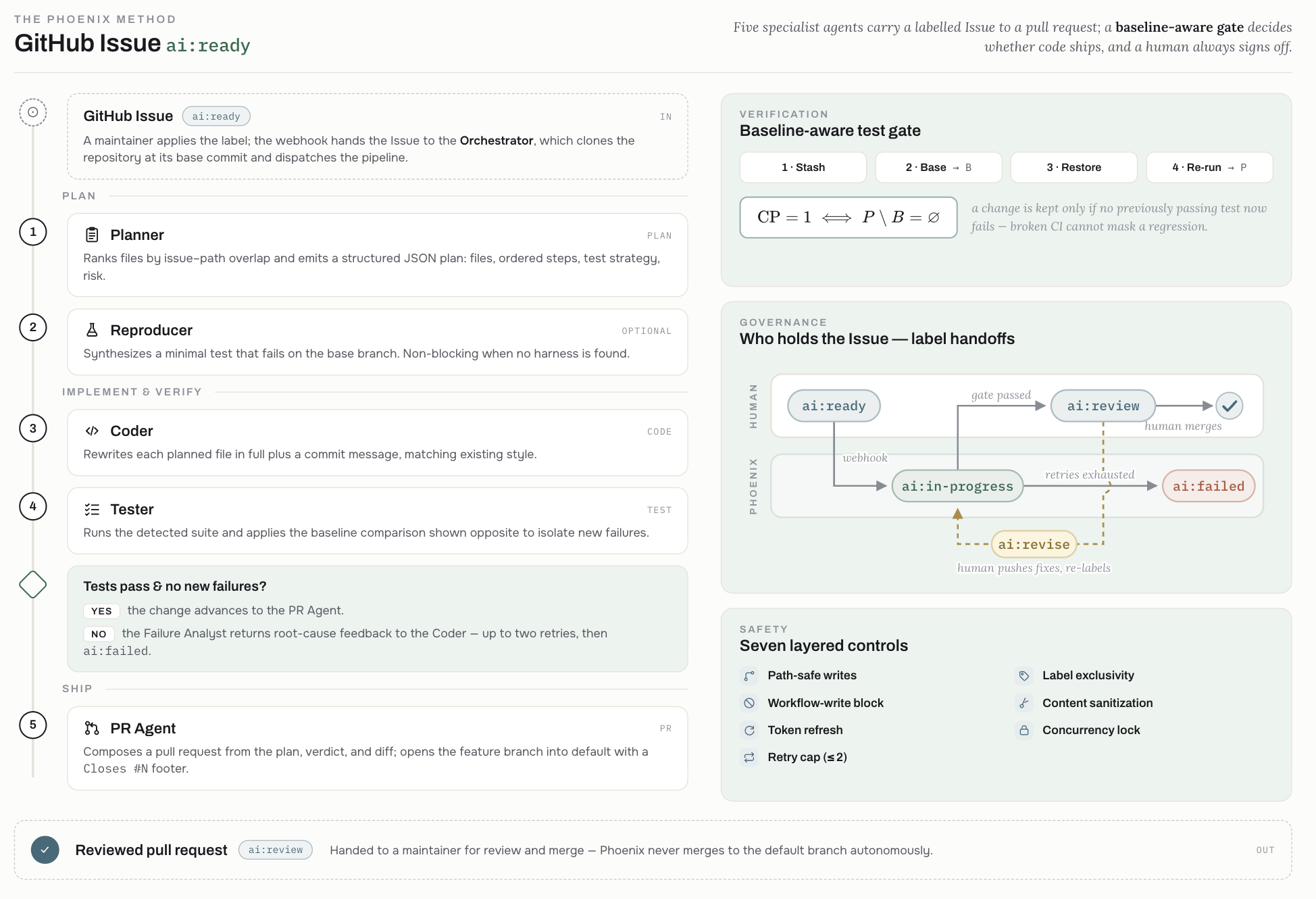}
    \captionof{figure}{Overview of the Phoenix pipeline. Six specialized agents carry a labeled GitHub issue from planning to a human-reviewed pull request, with a baseline-aware test gate. GitHub labels serve as the persistent state store, and seven layered safety controls constrain every run.}
    \label{fig:method}
\end{strip}

\section{Introduction}

Open-source software repositories on GitHub accumulate thousands of unresolved issues representing bugs, regressions, and feature gaps \cite{swebok2014}. Resolving even a modest fraction of these issues demands substantial engineering effort. A developer must understand the context of the codebase, reproduce the problem, identify the correct code change, implement it, write or update tests, and manage the review workflow. At scale, this creates a persistent maintenance bottleneck \cite{fowler2018refactoring}.

Large language models with strong code understanding \cite{chen2021codex,li2023starcoder,guo2024deepseek} have motivated systems that attempt to automate this process. Devin \cite{devin2024}, SWE-Agent \cite{yang2024sweagent}, and AutoCodeRover \cite{zhang2024autocoderover} apply LLM agents to GitHub issue resolution and report meaningful resolution rates on SWE-bench \cite{jimenez2024swebench}. However, optimizing for resolution rate alone can come at the cost of safety. In this regard, systems that resolve 30\% of issues but introduce regressions in another 20\% provide uncertain net value in production.

Phoenix instead prioritizes \emph{correctness-first} operation. Every change is gated behind automated tests, pre-existing failures are accounted for via baseline comparison, and every output is an auditable pull request for human review. Its safety mechanisms address concrete hazards observed in production deployment, notably content filtering by Application Programming Interface (API) gateways, authentication token expiry during long-running operations, Continuous Integration (CI) permission boundaries, and pre-existing broken test suites.

\noindent This paper makes the following contributions:
\begin{enumerate}
    \item \textbf{Architecture}. We propose an architecture with six-agent pipeline, including an optional Reproducer before coding, with configurable resolution mode, a label-driven state machine, seven layered safety controls, and an explicit design rationale for the decomposition.
    %
    %
    \item \textbf{Baseline-aware and Empirical Evaluation}. A test evaluation strategy that distinguishes pre-existing failures from Phoenix-introduced regressions, which enable assessment on repositories with broken CI pipelines.
    We also provide an empirical evaluation on SWE-bench Lite ($n{=}24$, $12$ repositories): $45.7\%$ oracle resolution, a $94.9\%$ oracle-eligible resolution on the production webhook path, and a $42$-issue pilot across $14$ repositories.

    \item \textbf{Deployment lessons}. We finally provide a characterization of the hazards encountered during production deployment and the mechanisms used to address them.
\end{enumerate}

\section{Background and Related Work}
\label{sec:related}

\subsection{LLM-Based Code Agents and SWE-bench}
\noindent SWE-bench \cite{jimenez2024swebench} established the standard benchmark for GitHub issue resolution. Each task is an issue description and repository state, with success measured by test passage. Recent systems with retrieval-augmented generation reach $30$--$40\%$ on the verified subset. SWE-Agent \cite{yang2024sweagent} introduced a structured agent-computer interface (ACI); AutoCodeRover \cite{zhang2024autocoderover} adds AST-based fault localization. However, both target offline benchmark performance.
Phoenix targets production deployment - real webhooks, real branches, real pull requests - and prioritizes correctness preservation over raw resolution rate.

\subsection{Automated Program Repair and Refactoring}
Automated Program Repair (APR) techniques \cite{goues2019apr}, genetic, semantic, and learning-based \cite{legoues2012genprog, nguyen2013semfix, tufano2019learning} 
%
%
target observed failures, notably failing tests. Phoenix targets requested changes based on issue description while preserving all passing tests, a complementary evaluation criterion. Similarly, refactoring tools \cite{fowler2018refactoring,tsantalis2018refactoringminer} apply predefined transformation templates. Phoenix interprets free-form issue descriptions without a fixed template library.

\subsection{Multi-Agent Systems and Tool Use}

Chain-of-thought prompting \cite{wei2022cot}, ReAct \cite{yao2023react}, and Toolformer \cite{schick2023toolformer} establish that LLMs benefit from explicit reasoning traces and tool invocation. Phoenix instantiates these ideas via multi-agent decomposition: each agent reasons within a narrow, fixed scope, and the pipeline provides multi-step execution scaffolding.

\subsection{AI Safety in Code Generation}

The risks of autonomous code modification in production are well documented. Untrusted Large Langue Models (LLMs) outputs can introduce security vulnerabilities \cite{pearce2022security}, violate invariants, or trigger non-obvious regressions. Phoenix addresses these risks structurally by enforcing that the system never merges code to the default branch autonomously. Instead, it always creates a pull request for human review, and enforces test verification before PR creation.

\section{System Architecture}
\label{sec:architecture}

\begin{figure}
    \centering
    \includegraphics[width=1\linewidth]{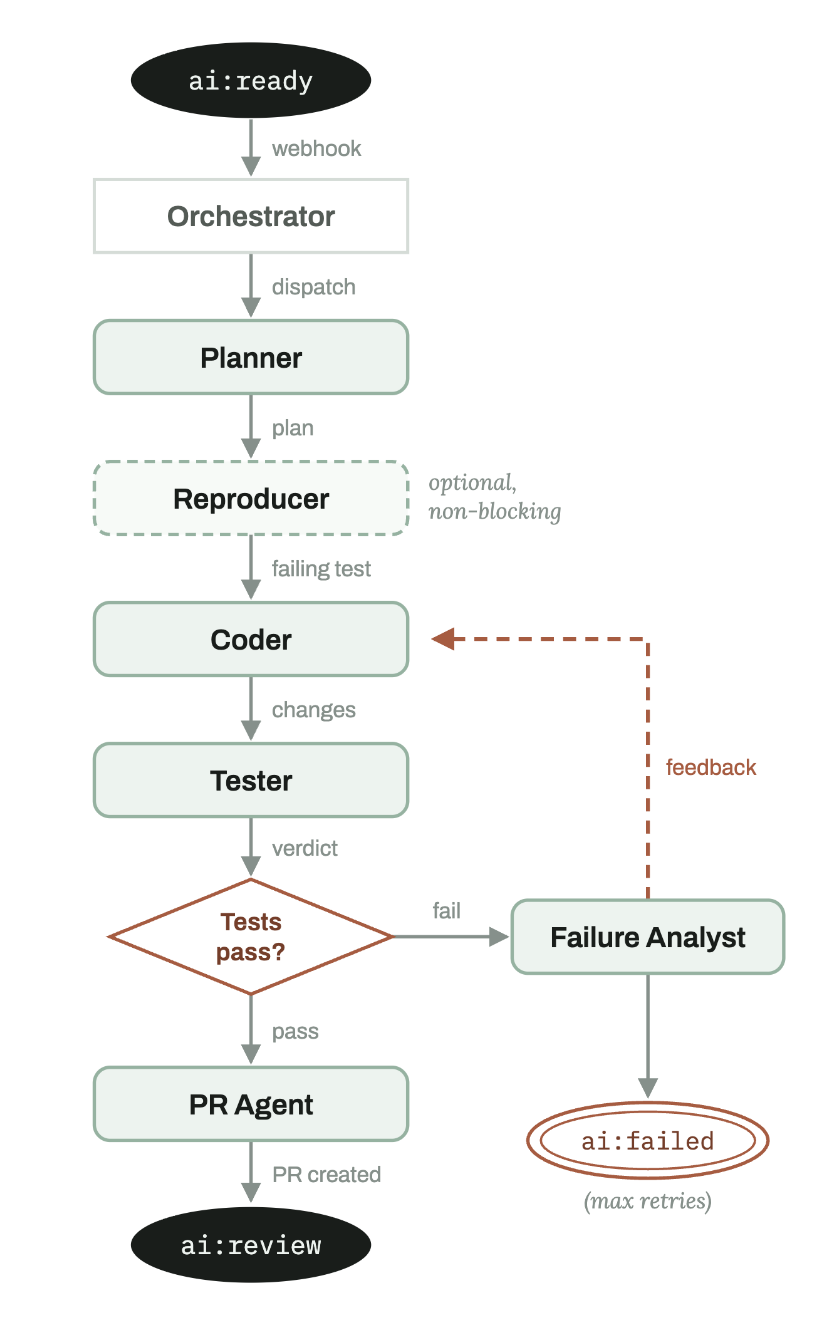}
\caption{Phoenix agent pipeline. After planning, the Reproducer (when enabled) attempts to add a failing test that demonstrates the issue on the base branch; implementation and testing follow, with Failure Analyst feedback on test failure. Reproducer failure is non-blocking. On retry exhaustion the issue is labeled \texttt{ai:failed}.}
\label{fig:architecture}
\end{figure}

Phoenix orchestrates six specialized LLM agents in a closed-loop pipeline, coordinated through a central Orchestrator. Fig. ~\ref{fig:method} provides the broad overview of the pipeline, and Fig ~\ref{fig:architecture} illustrates the end-to-end agentic flow.
%

\subsection{Design Rationale}
\label{sec:rationale}

The six-agent decomposition mirrors the stages of a conventional issue-resolution workflow-triage and planning, reproduction, implementation, testing, failure diagnosis, and review hand-off-so that each agent corresponds to one stage with a distinct input--output contract. This extends the staged structure of prior systems, such as the separation of fault localization from patch generation in AutoCodeRover \cite{zhang2024autocoderover}, and is consistent with evidence that LLMs perform more reliably under short, narrowly scoped instructions and explicit step-by-step scaffolding \cite{wei2022cot,yao2023react}. The decomposition also yields two engineering benefits. Each agent can be tested in isolation against fixed inputs, and failures can be attributed to a specific stage.
%

\subsection{Agent Roles and Responsibilities}

\subsubsection{Planner Agent}
The Planner receives the issue title and body sanitized for safety, repository file tree, most-relevant source file excerpts, recent issue comments, and any screenshot-derived visual context. It outputs a structured JSON plan: a one-sentence summary, high-level approach, files to modify and create, ordered implementation steps each with target file and action, test strategy, and risk level.
File relevance is determined by a scoring function that counts keyword overlap between the issue text and file path, preferring deeper files over shallow configuration files. For large repositories with over 500 source files, the maximum excerpt count and per-file character limit are reduced to keep prompt size within gateway limits. The Planner also detects the project language and instructs the Coder not to create files in other languages. Excerpts from the planned files are also supplied to the Reproducer when it runs.

\subsubsection{Reproducer Agent}
When enabled, the Reproducer runs immediately after planning. It receives the issue text, plan summary, and relevant file excerpts, and emits a small synthetic test that should fail on the unmodified base branch to demonstrate the reported bug. If it cannot produce a confirmed failing test after bounded attempts, the step is marked skipped and the pipeline continues; reproducer failure is by design non-blocking so that issues without an easy test harness still reach the Coder. On SWE-bench runs the Reproducer was enabled and never skipped, but the stricter ``reproducer test passes after the fix'' criterion is orthogonal to the SWE-bench oracle score.

\subsubsection{Coder Agent}
The Coder receives the plan, current contents of all plan-referenced files, and any prior failure feedback. It produces complete file contents for each changed file as a structured JSON object, including the commit message. The Coder's system prompt enforces a complete file content with no placeholders, adherence to existing code style, language-correct file types, and a self-verification step in which the Coder traces through its own test cases before finalizing output. If the raw output is not valid JSON, a single repair pass is attempted before failure.
All file writes are applied to the local clone and validated against path traversal before disk writes. Writes to \texttt{.github/workflows/} are blocked unconditionally.

\subsubsection{Tester Agent}
The Tester installs project dependencies and executes the test suite using the framework detected for the project. Test results are parsed to extract the set of failing test identifiers. A configurable resolution mode determines whether an issue counts as resolved when project tests pass, baseline comparison succeeds and the reproducer test passes after the fix. When tests fail, the Tester applies a baseline comparison strategy as follows.

\begin{figure}
    \centering
    \includegraphics[width=1\linewidth]{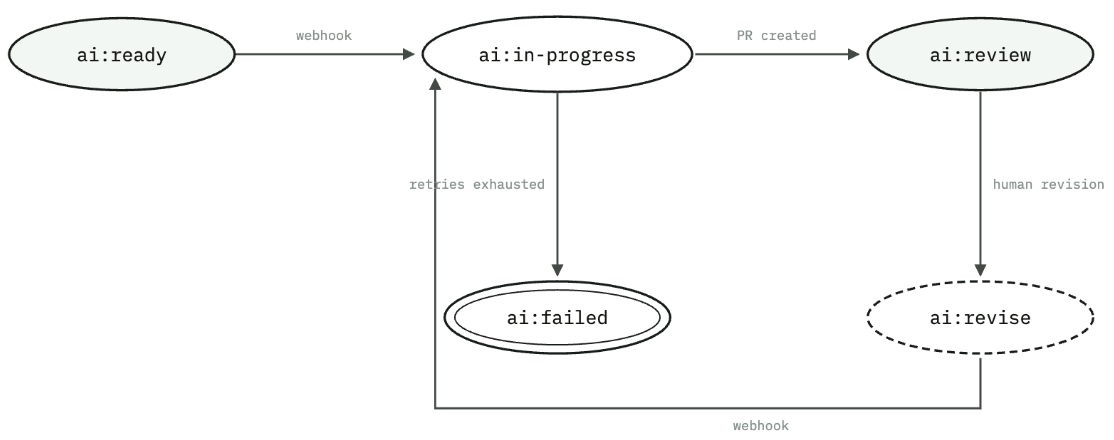}
\caption{Baseline-aware test evaluation. Phoenix stashes its changes and runs the suite on the unmodified branch (step 2) to collect baseline failures $B$, then restores changes and re-runs (step 4) to collect post-change failures $P$. Correctness is preserved iff $P \setminus B = \emptyset$: no previously passing test now fails.}
\label{fig:baseline}
\end{figure}

\begin{enumerate}
    \item Stash Phoenix's changes
    \item Run the test suite on the unmodified base branch.
    \item Pop the stash
    \item Then compute the set difference to obtain new failures count by dividing the number post-changes failures by the number baseline failures.
\end{enumerate}
If the baseline already fails and Phoenix introduced no new failures, the run is classified as \emph{correctness preserved}. This matters when evaluating Phoenix on repositories with broken CI pipelines. Fig.~\ref{fig:baseline} illustrates the strategy.

\subsubsection{Failure Analyst Agent}
When the Tester detects new failures, the Failure Analyst receives the full test output, the failure set, and a summary of the run. It produces structured feedback identifying likely root causes and specific suggestions for the Coder, which is appended to the Coder's context for the retry attempt. At most two retry cycles are permitted; on exhaustion, the issue is labeled \texttt{ai:failed} and a diagnostic comment is posted.

\subsubsection{PR Agent}
The PR Agent composes a pull-request title and body from the issue description, implementation plan, test verdict, and list of changed files. It opens the pull request from the feature branch to the default branch via the GitHub API, referencing the originating issue with a \texttt{closes\#{N}} footer, where $N$ is the issue number. The issue label is transitioned to \texttt{ai:review} to indicate the pull request is ready for human review.

\subsection{State Machine and Orchestration}

\begin{figure}[!htbp]
    \centering
    \includegraphics[width=1\linewidth]{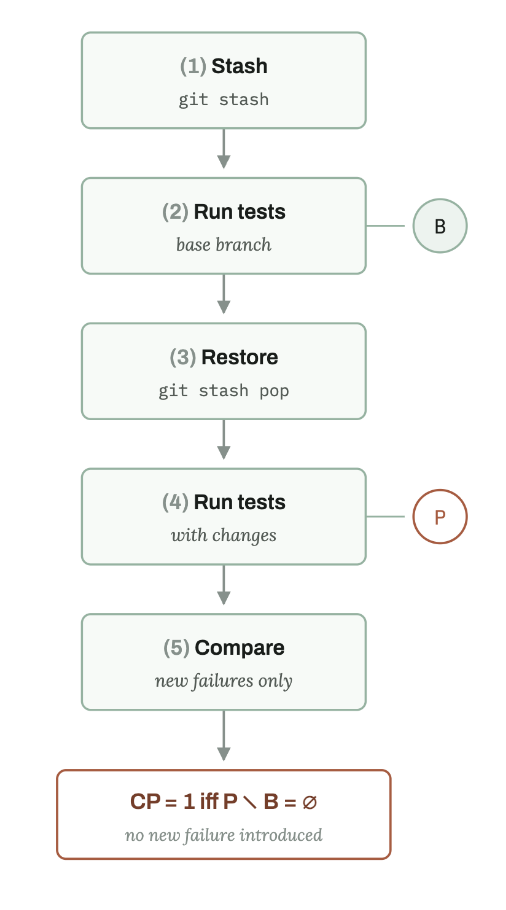}
\caption{GitHub label state machine. Labels serve as the persistent state store; transitions are atomic and mutually exclusive. The \texttt{ai:revise} loop enables iterative human-in-the-loop refinement. \texttt{ai:failed} is a terminal failure state; \texttt{ai:review} is the handoff to human review.}
\label{fig:statemachine}
\end{figure}

Phoenix maintains a per-issue state using GitHub labels as the persistent state store, without requiring a local database. Fig.~\ref{fig:statemachine} shows the five-state machine.

%

Label transitions are atomic. The new label replaces all AI-state labels. This ensures at most one active state per issue. A loop-prevention filter discards label events originating from the application installation itself. A per-installation lock serializes concurrent issues on the shared clone; separate repositories run in parallel.

\subsection{Safety Mechanisms}
\label{sec:safety}

Seven layered safety mechanisms were derived from failures encountered during development and deployment. We describe each with its motivating failure mode.

\textbf{1. Path-traversal prevention.} File writes validate that the resolved path does not escape the repository root. 
\textbf{2. Label-state exclusivity.} The transition function atomically removes all AI state labels before applying the new one, preventing stale label accumulation from interrupted runs.

\textbf{3. Workflow file guardrail.} The writes to  \texttt{.github/workflows/} directory are silently dropped. The GitHub application installation lacks the workflow push permission, and including such writes causes the entire push to be rejected.

\textbf{4. Content sanitization.} Issue bodies are truncated to $1,500$ chars, fenced code blocks replaced with one-line summaries, and traceback lines stripped before inclusion in any LLM prompt. Large stack traces and JSON payloads trigger WAF 403 responses from the LLM API gateway.

\textbf{5. Retry cycle limit.} At most two Failure Analyst feedback rounds are permitted; a no-progress detector terminates immediately if the second Coder attempt produces identical changes to the first.

\textbf{6. Concurrency serialization.} A per-installation lock ensures one issue is processed at a time, preventing conflicts on the shared local clone.

\textbf{7. Installation token refresh.} GitHub App tokens expire after one hour. The Orchestrator proactively refreshes tokens older than 50 minutes and updates the git remote URL before each pull.

A measurement caveat applies throughout. This is how often each mechanism triggered was reconstructed from operational logs after the fact rather than from instrumented counters, so any trigger frequencies we report should be read as lower bounds on how often the corresponding hazard occurred. 


\subsection{Implementation Details}

Phoenix is implemented in Python 3.11. LLM calls use LangChain \cite{chase2023langchain} with Claude Sonnet~4 \cite{anthropic2025claude} as the backbone model. It uses a $0.2$ temperature, with $8,192$-token context per call. Repository operations use GitPython; GitHub API interactions use PyGithub. The webhook server is built on FastAPI with a Server-Sent Events (SSE) endpoint for real-time run status streaming to the evaluation harness. The Reproducer and resolution modes are controlled via environment configuration, notably \texttt{USE\_REPRODUCER}, \texttt{RESOLUTION\_MODE}, etc... The system is packaged as a GitHub application installable via a single manifest file, requiring no per-repository configuration beyond the \texttt{ai:ready} label.

\section{Evaluation}
\label{sec:evaluation}

Our primary quantitative benchmark is SWE-bench Lite. Section~\ref{sec:pilot42} describes a complementary qualitative pilot on 42 real issues across 14 repositories.

\subsection{SWE-bench Lite}
\label{sec:swebench}

We evaluated Phoenix on SWE-bench Lite \cite{jimenez2024swebench}. We selected eight Python repositories supported by our harness - Astropy, Django, Flask, Matplotlib, pytest, Requests, scikit-learn, SymPy. We take three instances per repository from the Lite split. For each instance we fork the upstream repository at the dataset's base commit, mirror the issue on the fork, and trigger Phoenix via the \texttt{ai:ready} label on the production webhook path, exercising the full deployed pipeline. After Phoenix finishes, we apply the dataset's oracle test patch to the PR branch and run the official \texttt{FAIL\_TO\_PASS} and \texttt{PASS\_TO\_PASS} tests, following the SWE-bench protocol. A run counts as 
oracle-resolved when all \texttt{FAIL\_TO\_PASS} tests pass and no \texttt{PASS\_TO\_PASS} test regresses.

Table~\ref{tab:swebench} gives per-repository outcomes. Phoenix achieved oracle resolution on $18/24$ instances. On every oracle-passing run, correctness preservation held on the oracle's regression check. Six runs did not yield a passing oracle score. Five ended with terminal label \texttt{ai:failed} before a successful pipeline (twice on matplotlib, flask, pytest, scikit-learn), and one scikit-learn instance hit the evaluator's wall-clock wait cap of $45$ minutes while Phoenix was still in progress. Among runs with oracle pass, mean time from applying \texttt{ai:ready} to terminal label was $170s$.

\begin{table}[htbp]
\centering
\caption{SWE-bench Lite: oracle resolution by repository (3 instances each).}
\label{tab:swebench}
\begin{tabular}{lcc}
\toprule
\textbf{Repository} & \textbf{Oracle (pass/total)} & \textbf{Resolved} \\
\midrule
\texttt{astropy/astropy}       & 3/3 & 100\% \\
\texttt{django/django}         & 3/3 & 100\% \\
\texttt{pallets/flask}         & 2/3 & 67\% \\
\texttt{matplotlib/matplotlib} & 1/3 & 33\% \\
\texttt{pytest-dev/pytest}     & 2/3 & 67\% \\
\texttt{psf/requests}         & 3/3 & 100\% \\
\texttt{scikit-learn/scikit-learn} & 1/3 & 33\% \\
\texttt{sympy/sympy}           & 3/3 & 100\% \\
\midrule
\textbf{Total} & \textbf{18/24} & \textbf{75\%} \\
\bottomrule
\end{tabular}
\end{table}


\subsection{Pilot Evaluation on 42 Real Issues}
\label{sec:pilot42}

We additionally evaluate Phoenix on 42 real issues from 14 open-source repositories. For each, we forked the repo, mirrored 3 recent open issues, and applied \texttt{ai:ready} to trigger Phoenix. Issues span bug reports, edge-case fixes, and feature requests; no issues were simplified or annotated. Table~\ref{tab:repos} lists the repositories.

\begin{table}[htbp]
\centering
\caption{Evaluation Repository Set. Repository names link to the evaluation forks (\texttt{github.com/kkipngenokoech/\ldots}) where Phoenix's pull requests are publicly visible.}
\label{tab:repos}
\begin{tabular}{llll}
\toprule
\textbf{Repository} & \textbf{Language} & \textbf{Tier} & \textbf{Issues} \\
\midrule
\href{https://github.com/kkipngenokoech/sqlparse}{\texttt{andialbrecht/sqlparse}}     & Python     & easy   & 3 \\
\href{https://github.com/kkipngenokoech/requests}{\texttt{psf/requests}}              & Python     & easy   & 3 \\
\href{https://github.com/kkipngenokoech/axios}{\texttt{axios/axios}}               & JavaScript & easy   & 3 \\
\href{https://github.com/kkipngenokoech/gson}{\texttt{google/gson}}                & Java       & easy   & 3 \\
\midrule
\href{https://github.com/kkipngenokoech/marshmallow}{\texttt{marshmallow-code/marshmallow}} & Python  & medium & 3 \\
\href{https://github.com/kkipngenokoech/docusaurus}{\texttt{facebook/docusaurus}}        & TypeScript & medium & 3 \\
\href{https://github.com/kkipngenokoech/cal.com}{\texttt{calcom/cal.com}}             & TypeScript & medium & 3 \\
\href{https://github.com/kkipngenokoech/lucene}{\texttt{apache/lucene}}              & Java       & medium & 3 \\
\href{https://github.com/kkipngenokoech/langchain}{\texttt{langchain-ai/langchain}}     & Python     & medium & 3 \\
\midrule
\href{https://github.com/kkipngenokoech/pytest}{\texttt{pytest-dev/pytest}}          & Python     & hard   & 3 \\
\href{https://github.com/kkipngenokoech/scikit-learn}{\texttt{scikit-learn/scikit-learn}}  & Python     & hard   & 3 \\
\href{https://github.com/kkipngenokoech/core}{\texttt{vuejs/core}}                 & TypeScript & hard   & 3 \\
\href{https://github.com/kkipngenokoech/grafana}{\texttt{grafana/grafana}}            & TypeScript & hard   & 3 \\
\href{https://github.com/kkipngenokoech/keycloak}{\texttt{keycloak/keycloak}}          & Java       & hard   & 3 \\
\midrule
\textbf{Total}             &            &        & \textbf{42} \\
\bottomrule
\end{tabular}
\end{table}

We grade repositories into three tiers to test Phoenix across varying levels of difficulty. Easy repositories are well-maintained libraries with clear issue semantics, comprehensive test suites, and codebases below $50,000$ lines of code; examples include \texttt{sqlparse}, \texttt{requests}, and \texttt{axios}. Medium repositories involve multi-file concerns, cross-cutting API changes, or a combined frontend and backend structure, with codebases of roughly $50,000$ to $500,000$ lines; examples include \texttt{cal.com}, a full-stack Next.js application, and \texttt{langchain}, a large Python ecosystem. Hard repositories are production-grade systems of up to $1.4M$ lines with complex domain logic and heterogeneous file structures, where resolving an issue requires deep codebase understanding; examples include \texttt{grafana}, \texttt{scikit-learn}, and \texttt{pytest}.

\subsection{Evaluation Protocol}

For each issue, we measure \emph{correctness preservation} ($CP$): whether the test suite still passes after Phoenix's changes, accounting for pre-existing baseline failures. Formally:

\[
CP(i) = \begin{cases}
1 & \text{if post-changes tests pass, OR} \\
  & \text{baseline fails AND new failures} = \emptyset \\
0 & \text{otherwise}
\end{cases}
\]

For Java repositories without local build tools, $CP(i) = 1$ by inspection of the generated code (no build-time error detected), deferring runtime verification to the PR review stage.


\subsection{Results}
\label{sec:pilotresults}

Table~\ref{tab:results} summarizes the pilot results. Phoenix achieved 100\% correctness preservation on all 42 issues; we emphasize that $CP$ measures the absence of regressions, not functional adequacy.

\begin{table}[htbp]
\centering
\caption{Phoenix Evaluation Results by Difficulty}
\label{tab:results}
\begin{tabular}{lrrr}
\toprule
\textbf{Tier} & \textbf{Issues} & \textbf{CP (pass/total)} & \textbf{Avg time (s)} \\
\midrule
Easy   & 12 & 12/12 (100\%) & - \\
Medium & 15 & 15/15 (100\%) & - \\
Hard   & 15 & 15/15 (100\%) & 122 \\
\midrule
\textbf{Total} & \textbf{42} & \textbf{42/42 (100\%)} & \\
\bottomrule
\end{tabular}
\end{table}

On hard-difficulty repositories, the mean end-to-end resolution time, which measures from the webhook receipt to PR creation was $122$ seconds, the minimum time was $68s$, and the maximum time $198s$. Timing data for easy and medium tiers was not recorded in the run artifacts; empirically these tiers completed within $2$--$4$ minutes per issue. All 14 repositories achieved 3/3 correctness preservation on $CP$.

\subsection{Pilot observations}

Three observations stand out from the pilot runs. First, 10 of the 11 non-Java repositories had pre-existing test failures, so baseline comparison was necessary to avoid falsely attributing those failures to Phoenix. Second, runs on large codebases such as \texttt{grafana} and \texttt{scikit-learn} depended on the Planner's keyword-based file ranking, which succeeds only when terms from the issue text match file paths. Third, repositories with oversized issue bodies, such as \texttt{langchain}, required content sanitization to avoid blocking by the API gateway firewall.

Manual inspection of the 42 pilot pull requests reveals two distinct quality tiers. Roughly half make targeted edits to real modules. The other half place generic scaffolding under invented paths, typically \texttt{src/core/config.py}, which preserves correctness under $CP$ without fixing the reported bug.

\section{Discussion}
\label{sec:discussion}

\subsection{What the Metrics Do and Do Not Measure}
\label{sec:reprodiscuss}

Correctness preservation, our primary pilot metric, is a necessary but not sufficient condition for resolving an issue. As reported in section~\ref{sec:pilot42}, roughly half of the generated pull requests modify existing repository files, while the other half place code at paths that did not previously exist in the repository. Both halves pass $CP$ because neither introduces new test failures, but only the first represents a genuine fix. The 100\% $CP$ rate should therefore be read as evidence that Phoenix introduced no regressions, not that it resolved all 42 issues. The gap is driven by localization rather than code synthesis. The Planner ranks files by lexical overlap between issue text and file paths; when an issue describes a bug in vocabulary that appears in no file name, which is common for behavioral bugs, the ranker returns nothing and the Planner falls back to inventing generic paths. Consistent with this explanation, the well-targeted pull requests concentrate in repositories whose module names mirror their domain vocabulary, such as \texttt{requests/auth.py} for an authentication issue. Semantic retrieval over the repository structure directly targets this failure mode and is our highest-priority future work. We nonetheless retain $CP$ as the primary metric because it is objectively measurable and safety-aligned: a system that causes regressions is more harmful than one that does nothing, and the \texttt{ai:review} label makes explicit that functional adequacy still requires human confirmation.

\subsection{Lessons from Deployment}

The seven safety mechanisms were not designed speculatively but added in response to a specific failure observed during development or evaluation, from gateway rejection of oversized issue bodies to token expiry during long runs. Our experience suggests that safety in deployed code-modification systems benefits from empirical safety engineering, namely running the system, observing its failure modes, and adding targeted defenses, in addition to a priori design.
The six-agent decomposition delivered its expected engineering benefits in practice: agent-level unit tests against fixed inputs caught prompt regressions, and pipeline failures were attributable to a specific stage from logs alone. The optional Reproducer adds latency but, when successful, grounds the implementation in a concrete failing test.

\subsection{Limitations}

Our evaluation and design carry three limitations. First, file localization is lexical: the Planner's ranker matches issue vocabulary against file names, and when no match exists it produces pull requests at invented paths. Roughly half of the pilot pull requests fall into this category, the capped file excerpts aggravate the problem on very large repositories, and $CP$ cannot distinguish these from genuine fixes. Second, the verification signal is imperfect. The test suite is the sole automated correctness check, so low coverage may mask regressions and flaky tests may produce false failures; for Java repositories, where Maven and Gradle are unavailable in our environment, $CP$ rests on code inspection rather than test execution. In all cases, passing tests does not imply the issue is resolved, and human review of each pull request remains necessary. Third, we did not compare against a single-agent baseline; the decomposition is motivated by prior work and engineering considerations.

\section{Future Work}
\label{sec:future}

The highest-priority improvement is \textbf{semantic retrieval} over the repository AST to replace the keyword-based file ranker, directly addressing the file-path targeting failures identified in Sections~\ref{sec:pilot42} and~\ref{sec:discussion}. Complementary directions include a \textbf{containerized build environments} per run to enable full test execution for Java, C/C++, and Rust without host toolchain dependencies; a \textbf{functional adequacy evaluation} via domain-expert PR review to quantify the gap between correctness preservation and true resolution rate; a \textbf{Security Analyst agent} invoking static analysis tools before PR creation; and a \textbf{per-repository memory module} to accumulate coding conventions across runs. A stress-tier evaluation on repositories above $1M$ LOC would characterize the limits of Phoenix's context management and validate a retrieval-augmented variant.
We reported a curated $n{=}24$ slice of Lite under the production webhook protocol. For fair comparison with prior work \cite{yang2024sweagent,zhang2024autocoderover,jimenez2024swebench}, future work should run Phoenix through the official SWE-bench Lite or Verified harness, reporting the percentage resolved on the full split alongside existing leaderboard entries, and should additionally evaluate strong off-the-shelf baselines on the same instances under the same protocol, to isolate the contribution of the multi-agent design.

\section{Conclusion}

We presented Phoenix, a multi-agent LLM system for safe, end-to-end GitHub issue resolution with six specialized agents, including an optional Reproducer before implementation. On SWE-bench Lite, Phoenix achieves a 75\% oracle resolution rate, with no pass-to-pass regressions on oracle successes; we caution that this slice is not protocol-matched to published leaderboard results. On a separate 42-issue pilot across 14 repositories, Phoenix achieves $100\%$ correctness preservation, with a mean resolution time of $122$ seconds on hard repositories; manual inspection shows roughly half of those PRs are well-targeted fixes, with the remainder limited by Planner file localization. The contributions are: a six-agent pipeline with an explicit design rationale and configurable resolution modes; baseline-aware testing; SWE-bench Lite numbers with explicit comparability caveats; and seven safety mechanisms derived from deployment experience. Our experience suggests that reliable autonomous code modification in production requires treating safety as a first-class engineering requirement: each of Phoenix's seven safety mechanisms was added in response to a concrete failure mode encountered during deployment, and we expect other deployed systems to encounter similar failure modes. Semantic retrieval remains the primary direction for addressing the Planner's localization limits. Phoenix is available as an open-source GitHub App, deployable on any repository with a single manifest installation.

\section*{Acknowledgments}
The authors thank the open-source maintainers of all 14 evaluation repositories for their publicly available codebases and issue trackers.

\section*{Code Availability}

Source code, evaluation scripts, and result artifacts are publicly available at \url{https://github.com/kkipngenokoech/phoenix}. All 42 evaluation pull requests are publicly visible on the fork repositories listed in Table~\ref{tab:repos}. Each fork is accessible at \texttt{github.com/kkipngenokoech/}; Phoenix PRs are labeled \texttt{ai:review} and can be browsed under the Pull Requests tab of each fork.

\bibliographystyle{ieeetr}
\bibliography{references}

\end{document}